\documentstyle[preprint,prd,aps]{revtex}\tighten

\newcommand{\beq}{\begin{equation}}
\newcommand{\beqa}{\begin{eqnarray}}
\newcommand{\eeq}{\end{equation}}
\newcommand{\eeqa}{\end{eqnarray}}
\newcommand{\vphi}{\varphi}

\begin{document}
\widetext
\draft
\preprint{\tighten\vbox{\hbox{UTAP-360}\hbox{gr-qc/0001029}}}

\title{
Scalar-Tensor Gravity in Two 3-brane System
}

\author{Takeshi Chiba\footnote{
Present Address: Department of Physics, Kyoto University, Kyoto
606-8502, Japan.}
}

\address{
Department of Physics, University of Tokyo, 
Tokyo 113-0033, Japan
}
\date{\today}
\maketitle

\bigskip

\begin{abstract}
We derive the low-energy effective action of four-dimensional gravity
in the Randall-Sundrum scenario in which two 3-branes of opposite 
tension reside in a five-dimensional spacetime. 
The dimensional reduction with the Ansatz for the radion field by 
Charmousis et al., which solves five-dimensional linearized field 
equations,  results in a class of scalar-tensor gravity theories. 
In the limit of vanishing radion fluctuations, the effective action  
reduces to the Brans-Dicke gravity in accord with the results of 
Garriga and Tanaka:  Brans-Dicke gravity with the corresponding
Brans-Dicke parameter  $0< \omega < \infty$ (for positive tension
brane) and $-3/2< \omega <0$ (for negative tension brane). 
In general the gravity induced a brane belongs to a class of
scalar-tensor gravity with the Brans-Dicke parameter which is 
a function of the interval and the radion. In particular, gravity on 
a positive tension brane contains an attractor mechanism toward the 
Einstein gravity.
\end{abstract}


\vfill\eject

\section{introduction}

Randall and Sundrum recently proposed a mechanism to solve the
hierarchy problem \cite{rs1} using two 3-branes of opposite tension
residing embedded in a five dimensional spacetime. In this scenario we 
are assumed to live in the negative tension brane. 
They also found that even the graviton is trapped on the positive
tension brane in the same set up \cite{rs2} (see also \cite{ver}).

It is natural to ask how gravity look like on the brane. 
In a recent paper, Garriga and Tanaka have analyzed the metric
perturbation equation on the 3-brane in the background spacetime of
Randall and Sundrum and have shown that gravity on the brane is
described by Brans-Dicke gravity \cite{gt} (for a single 3-brane, see
\cite{sms}). Charmousis et al. recently proposed an Ansatz for
the radion which solves five-dimensional linearized equations of
motion \cite{cgr}. 

The purpose of this note is to provide alternative considerations on
the brane gravity based on the derivation of the low-energy effective
action on each 3-brane from the five dimensional action {\it \`a la} 
Kaluza-Klein. 
We take into account the degrees of freedom of the modulus field as
well as four-dimensional graviton fluctuations following the metric
Ansatz by Charmousis et al. We will see that 
gravity on the brane is actually a more general scalar-tensor gravity
theory, where the Brans-Dicke parameter is a function of the
Brans-Dicke scalar field, rather than the Brans-Dicke gravity, where
the Brans-Dicke parameter is a constant. 
The modulus field (the radion) plays the role of the Brans-Dicke scalar
field. We believe that our approach is technically simpler. The
calculations required in our approach are reduced considerably, and it
is easy to generalize to include matter fields.   
We show that the cosmological attractor mechanism toward the Einstein 
gravity is realized in  gravity on a positive tension brane. We also
make a brief comment on the cosmological constant problem in the
brane-world view of the universe. 

\section{Reduction in two three-brane system}

\subsection{Background}

We briefly review the background metric found by Randall and Sundrum 
\cite{rs1} to introduce our notations. The set up is that the 
five-dimensional Einstein gravity
with a cosmological constant with two 3-branes located in the orbifold
$S^1/{\bf Z}_2$ at the fifth dimensional coordinate $z=0$(positive
tension brane), $z=r_c$(negative tension brane). The action is
\beq
S=2\int d^4x\int_{0}^{r_c} d\phi\sqrt{-g_5}\left[
{M^3\over 2}R_5-2\Lambda\right]-\sigma_{(+)}\int d^4x\sqrt{-g_{(+)}}
-\sigma_{(-)}\int d^4x\sqrt{-g_{(-)}},
\eeq
where $M$ is the five-dimensional Planck mass defined in term of the 
five-dimensional gravitational constant $G_5$ as $M^{-3}=8\pi G_5$. 
$\sigma_{(\pm)}$ and $g_{(\pm)}$ are the brane tension and the
induced metric on the brane, respectively. 
Randall and Sundrum have shown that there exists a solution that
respects four-dimensional Poincar\'e invariance \cite{rs1}:
\beq
ds^2=e^{-2k |z|}\eta_{\mu\nu}dx^{\mu}dx^{\nu}+dz^2,
\label{back}
\eeq
only if $\Lambda, \sigma_{(+)}$ and  $\sigma_{(-)}$ are related as
\beq
\Lambda=-3M^3k^2, \sigma_{(+)}=-\sigma_{(-)}=3M^3k.
\label{relation}
\eeq

\subsection{Naive Ansatz}

We shall derive the four-dimensional low-energy effective action on a 
3-brane. In order to include massless
gravitational degrees of freedom (the zero modes about the background
spacetime Eq.(\ref{back})), we replace the Minkowski metric
$\eta_{\mu\nu}$ with a general metric $\bar{g}_{\mu\nu}(x)$ and 
$z$-direction length $r_c$ with a modulus field $T(x)$ \cite{rs1,gw}: 
\beq
ds^2=e^{-2kT(x)|z|}
\bar{g}_{\mu\nu}(x)dx^{\mu}dx^{\nu}+T(x)^2dz^2.
\label{metricnaive}
\eeq
Although this metric ansatz does not correctly describe the linearized
dynamics of massless fields, we present the analysis here because 
the results nevertheless seem simple and attractive.
The induced metric on the positive (or negative) tension brane is thus  
$g_{(+)\mu\nu}=\bar{g}_{\mu\nu}$ (or $ g_{(-)\mu\nu}= e^{-2
kr_cT}\bar{g}_{\mu\nu}$), respectively. 
Since we impose ${\bf Z}_2$ symmetry ($z\leftrightarrow -z$), 
massless vector fluctuations associated with the off-diagonal part of 
the metric are absent \cite{rs1,gw}.

\subsubsection{Effective Four-dimensional Action}

We can perform the $z$ integral to obtain the four-dimensional
action on the positive tension brane described by $g_{(+)\mu\nu}$:
\beqa
S_{(+)}&=&2\int d^4x\int_{0}^{r_c} dz\sqrt{-g_5}\left[
{M^3\over 2}R_5-2\Lambda\right]-\sigma_{(+)}\int
d^4x\sqrt{-g_{(+)}}-\sigma_{(-)}\int d^4x\sqrt{-g_{(-)}}
\nonumber\\
&=&2\int d^4x\int_{0}^{r_c} {dz}\sqrt{-g_{(+)}}e^{-4kTz}T
\left[{M^3\over 2}
e^{2kTz}\left(R_{(+)}-6e^{kTz}\Box_{(+)}e^{-kTz}\right)
-2\Lambda\right]\nonumber\\
&&-\sigma_{(+)}\int d^4x\sqrt{-g_{(+)}}-\sigma_{(-)}\int
d^4x\sqrt{-g_{(+)}}e^{-4 kr_cT}\nonumber\\
&=&\int d^4x\sqrt{-g_{(+)}}\left[{M^3(1-e^{-2 kr_cT})\over 2k}R_{(+)}-
3kr_c^2M^3e^{-2kr_cT}\left(\nabla_{(+)}T\right)^2\right.\nonumber\\
&&\left.-{(1-e^{-4kr_cT})\over k}\Lambda
-\sigma_{(+)}-\sigma_{(-)}e^{-4kr_cT}\right]\nonumber\\
&=&\int d^4x\sqrt{-g_{(+)}}\left[{M^3(1-e^{-2kr_cT})\over
2k}R_{(+)}-3kr_c^2M^3e^{-2kr_cT}\left(\nabla_{(+)}T\right)^2\right],
\label{4daction-posi-naive}
\eeqa
where we have used the relation Eq.(\ref{relation}) in the last equation.

The case of the negative tension brane is obtained by  the conformal
transformation such that $g_{(-)\mu\nu}=e^{-2
kr_cT}\bar{g}_{\mu\nu}=e^{-2kr_cT}g_{(+)\mu\nu}$, and the result is
\beq
S_{(-)}=\int d^4x\sqrt{-g_{(-)}}\left[{M^3(e^{2kr_cT}-1)\over
2k}R_{(-)}+3kr_c^2M^3e^{2kr_cT}\left(\nabla_{(-)}T\right)^2\right].
\label{4daction-nega-naive}
\eeq
The above result is also easily obtained by the change of signature such
that $k \rightarrow -k$, the meaning of which may be intuitively
clear: exchange of the locations of the branes. 

\subsubsection{Brans-Dicke parameter}

The action of the scalar-tensor gravity theory is given by \cite{will}
\beq
S=\int d^4x\sqrt{-g}{1\over
16\pi}\left(\Phi_{BD}R-{\omega(\Phi_{BD})\over \Phi_{BD}}(\nabla
\Phi_{BD})^2\right).
\eeq
The correspondence of the gravity theory on the brane to the
Brans-Dicke gravity is then immediate. The Brans-Dicke scalar field
 (or the inverse of the effective gravitational constant) is 
\beq
{1\over G_{\rm eff}^{(\pm)}}=\Phi_{BD}^{(\pm)}=
{16\pi M^3\over k}e^{\mp kr_cT}\sinh(
kr_cT)={2\over kG_5}e^{\mp kr_cT}\sinh(kr_cT),
\eeq
while the corresponding Brans-Dicke function is
\beq
\omega_{(\pm)}(T)=\pm 3e^{\pm kr_cT}\sinh(kr_cT).
\eeq
Remarkably, despite the defect of the metric Ansatz, both expressions 
are in perfect agreement with those of
Garriga and Tanaka\cite{gt} if the length parameter between the branes
is replaced with the modulus field $r_cT(x)$: the Brans-Dicke gravity
with the corresponding Brans-Dicke 
parameter  $0< \omega < \infty$ (for positive tension brane) and
$-3/2< \omega <0$ (for negative tension brane).

\subsection{CGR Ansatz}

Although the previous analysis based on the naive Ansatz seems simple and
attractive,  we seek after another metric Ansatz which does
solve the linearized equations of motion. 
As an example, we adopt the metric Ansatz proposed by
Charmousis et al.(CGR)\cite{cgr} (we shall consider the 
region $0<z<r_{c}$; the other region by 
the orbifold symmetry):
\beqa
&&ds^2=e^{-2kh(x,z)}
\bar{g}_{\mu\nu}(x)dx^{\mu}dx^{\nu}+h_{,z}^2dz^2,\label{metric}\\
&&h(x,z)=z+f(x)e^{2kz}.\label{ansatz}
\eeqa
The induced metric on each brane is thus respectively 
\beqa
&&g_{(+)\mu\nu}=e^{-2kf}\bar{g}_{\mu\nu},\\
&& g_{(-)\mu\nu}=\alpha^{-1}e^{-2\alpha kf}\bar{g}_{\mu\nu}.
\eeqa
Here we have introduced the notation $\alpha\equiv e^{2kr_c}$ for
convenience.  

\subsubsection{Effective Four-dimensional Action}

We then perform the $z$ integral to obtain the four-dimensional
action on the positive tension brane in terms of ${g}_{(+)\mu\nu}$:
\beq
S_{(+)}=
\int d^4x\sqrt{-g_{(+)}}\left[{M^3\over 2k}\left(1-{1\over \alpha}
e^{-2(\alpha -1) kf}\right)R_{(+)}-
3kM^3\left[\left(\alpha + {1\over \alpha}\right)e^{-2(\alpha -1)kf}
-2\right]\left(\nabla_{(+)}f\right)^2\right],
\label{4daction-posi}
\eeq
where we have again used the relation Eq.(\ref{relation}).

The case of the negative tension brane is obtained by  the conformal
transformation such that $g_{(-)\mu\nu}=\alpha^{-1}e^{-2\alpha
kf}\bar{g}_{\mu\nu}=\alpha^{-1}e^{-2(\alpha -1)kf}g_{(+)\mu\nu}$,
 and the result is
\beq
S_{(-)}=\int d^4x\sqrt{-g_{(-)}}\left[{M^3\over 2k}\left(\alpha
e^{2(\alpha -1)kf}-1\right)R_{(-)}+
3\alpha^2kM^3\left[\left(\alpha + {1\over \alpha}\right)e^{2(\alpha -1)kf}
-2\right]\left(\nabla_{(-)}f\right)^2\right].
\label{4daction-nega}
\eeq

\subsubsection{Brans-Dicke parameter}

The Brans-Dicke scalar field
 (or the inverse of the effective gravitational constant) is 
\beq
{1\over G_{\rm eff}^{(\pm)}}=\Phi_{BD}^{(\pm)}=
{2\over kG_5}e^{\mp k[(\alpha -1)f+r_c]}\sinh k[(\alpha -1)f+r_c]
\eeq
while the corresponding Brans-Dicke function is
\beq
\omega_{(\pm)}(f)=\pm 3e^{\pm k[(\alpha -1)f+r_c]}\sinh k[(\alpha
-1)f+r_c]
\left(1\mp {e^{\pm (\alpha -1)kf}\sinh [(\alpha -1)kf]\over \sinh^2(kr_c)}
\right).
\eeq
Both expressions are in  agreement with those of
Garriga and Tanaka\cite{gt} in the limit where the radion fluctuations 
are vanishing, $f \rightarrow 0$. 

Now let us consider what kind of gravity is induced on the brane if
the radion fluctuations are present. As long as $e^{2\alpha kf}\ll
\alpha \equiv e^{2kr_c}$ the metric ansatz Eq.(\ref{metric}) and 
Eq.(\ref{ansatz}) correctly describes the linearized dynamics of
massless fields.  For the positive tension brane, gravity on the brane 
is a class of scalar-tensor gravity theories with $\omega >0$ (for 
$e^{2\alpha kf}\ll \alpha$ we have $\omega= 3\alpha e^{2\alpha
kf}/2$) and can satisfy the constraint by the solar system
experiment($|\omega|>3000$ \cite{will2}) if $kr_c> 4$. 
In the limit $kr_c \rightarrow \infty$, the Einstein gravity is
recovered with the gravitational constant $G_4=kG_5$. 

On the other hand, gravity on the negative tension brane is a 
class of scalar tensor theories with 
$ \omega =-3/2 +9\alpha^{-1}e^{2\alpha kf}/2 <0$ (in the lowest order 
of ${\cal O}( \alpha^{-1}e^{2\alpha kf})$) which does not satisfy the 
constraint by the solar system experiment.
This does not immediately mean that the scenario of \cite{rs1} is not 
valid because we do not include massive degrees of freedom that may
give rise to a stabilizing potential for the modulus(see \cite{gw2,csaki}).

Moreover, gravity on the negative tension brane exhibits some peculiar 
behavior in the limit $kr_c \rightarrow \infty$: $G_{\rm eff}
\rightarrow 0$ and  $\omega \rightarrow -3/2$.  The vanishing of the 
gravitational constant would imply the absence of gravity force. 
Furthermore, $\omega=-3/2$ means that the scalar field degrees of
freedom is completely absorbed by the conformal transformation; 
the action is reduced to that of {\it vacuum} Einstein gravity 
(no scalar field) by the conformal transformation such that 
$g_{\mu\nu}^E=[\alpha e^{2(\alpha -1)kf}-1]g_{(-)\mu\nu}$.\footnote{We
note that in the limit  $kr_c \rightarrow \infty$ the Einstein 
conformal frame metric $g_{\mu\nu}^E$ coincides with the metric on the 
positive tension brane $g_{(+)\mu\nu}$.}
 In this sense, the scalar field degrees of freedom is also frozen.

\subsubsection{Attractor Mechanism}
 
The Einstein gravity is generically a cosmological attractor in
scalar-tensor theories (attractor mechanism \cite{dn}). Therefore, 
as far as gravity is concerned, the dilaton stabilization is not
always necessary. It is interesting to examine whether gravity on the
positive tension brane contains  a similar attractor mechanism 
although it requires slight abuse  the action 
Eq.(\ref{4daction-posi}) beyond the weak field approximation. 
To do so, we assume matter on the positive tension brane and 
shall work in the Einstein conformal frame defined by 
$g^E_{\mu\nu}=[1-\alpha^{-1}e^{-2(\alpha -1)kf}]g_{(+)\mu\nu}\equiv 
e^{-2a(\vphi)}g_{(+)\mu\nu}, 
2({da/ \kappa d\vphi})^2=(2\omega(f)+3)^{-1}$ 
so that the action Eq.(\ref{4daction-posi}) reduces to
\beq
S_{(+)}=\int d^4x \sqrt{-g_E}\left[{1\over 2\kappa^2}R_E-
{1\over 2}(\nabla_E \vphi)^2\right],
\eeq
where $\kappa^2=8\pi G_4\equiv 8\pi kG_5$. An important point of the 
attractor mechanism by Damour and Nordvedt \cite{dn} is that the 
function $a(\vphi)$ determines the cosmological dynamics of the
dilaton $\vphi$ and the minimum of it is the Einstein gravity
($a(\vphi)_{,\vphi}=0 \rightarrow \omega = \infty$). Therefore it is
sufficient to see the shape of $a(\vphi)$. For $e^{2\alpha kf}\ll
\alpha$, we find that $2a(\vphi)=-\ln(1-2\kappa^2\vphi^2/3)$ with 
$2\kappa^2\vphi^2=3\alpha^{-1}e^{-2\alpha kf}$; that is, $a(\vphi)$ 
indeed has a minimum at $\vphi =0$, which
corresponds to $\omega \rightarrow \infty$.\footnote{The attractor
  mechanism is also realized in the action Eq.(\ref{4daction-posi-naive}).}

\subsubsection{Adjusting Mechanism?}

Finally let us consider the situation where the bulk cosmological
constant $\Lambda$ is slightly perturbed from the assumed value 
Eq.(\ref{relation}), $\Lambda \rightarrow \Lambda + \delta \Lambda$,
while the brane tensions are fixed. Such a mismatch would induce an 
effective potential for the modulus $f(x)$ and give rise to the
dynamics of $f(x)$. However, because the modulus $f(x)$ couples
non-minimally to the curvature, the effective potential for $f(x)$ is 
given by performing the conformal transformation to the canonical
Einstein frame $g_{\mu\nu}^E= [1-\alpha^{-1}e^{-2(\alpha
-1)kf}]g_{(+)\mu\nu}$ (for simplicity we consider the positive tension 
brane), and it is given by \footnote{The potential of this type has been 
  discussed in the context of inflation\cite{guendelman}.}
\beq
V_{\rm eff}(f)= {\delta \Lambda\over k}\coth k[(\alpha -1)f+r_c].
\eeq
Since the minimum of the potential is not zero, there is no dynamical 
mechanism to enforce $V_{\rm eff}=0$. 
Hence,  contrary to the argument in \cite{st}, brane-world picture of
the universe seems to be compatible with a nonzero four-dimensional
cosmological constant (similar observations are made in \cite{dfgk,knk}): 
the four-dimensional background metric can be Minkowski
spacetime, de Sitter spacetime, or anti-de Sitter spacetime, depending
both on the bulk cosmological constant and on the brane tensions. 
The cosmological constant problem (in the present context the
fine-tuning between the bulk five-dimensional cosmological constant
and the brane tension) remains the problem.\footnote{For stability of
 the classical Minkowski background, see \cite{vv}.} 


\section{summary}

We have given  simple considerations on the brane gravity based on the
dimensional reduction. We have shown that gravity on the brane belongs
to a class of scalar-tensor gravity theory where the radion field plays
the role of the Brans-Dicke scalar field.
We have pointed out the possibility of the cosmological attractor
mechanism on the positive tension brane  
and have also made a brief comment on the cosmological
constant problem in the brane world scenario. 

\acknowledgments
We would like to thank Yasunori Nomura, Tomohiko Takahashi  for useful 
discussions and Takahiro Tanaka for
useful correspondence and comments. 
This work was supported in part by the JSPS under Grant No. 3596.

\end{document}